\documentclass{article}

\usepackage[dblblindworkshop,final]{neurips_2025}
\workshoptitle{The First Workshop on Generative and Protective AI for Content Creation}

\usepackage[utf8]{inputenc}
\usepackage[T1]{fontenc}
\usepackage{hyperref}
\usepackage{url}
\usepackage{booktabs}
\usepackage{amsfonts}
\usepackage{amsmath}
\usepackage{nicefrac}
\usepackage{microtype}
\usepackage{xcolor}
\usepackage{graphicx}
\usepackage{algorithm}
\usepackage{algorithmic}

\title{Generative AI Agents for Controllable and Protected Content Creation}

\author{
  Haris Khan \\
  National University of Sciences and Technology (NUST) \\
  Islamabad, Pakistan \\
  \texttt{mhariskhan.ee44ceme@student.nust.edu.pk}
  \And
  Sadia Asif \\
  Rensselaer Polytechnic Institute (RPI) \\
  New York, United States \\
  \texttt{asifs@rpi.edu}
}

\begin{document}

\maketitle

\begin{abstract}
The proliferation of generative AI has transformed creative workflows, yet current systems face critical challenges in controllability and content protection. We propose a novel multi-agent framework that addresses both limitations through specialized agent roles and integrated watermarking mechanisms. Unlike existing multi-agent systems focused solely on generation quality, our approach uniquely combines controllable content synthesis with provenance protection during the generation process itself. The framework orchestrates Director/Planner, Generator, Reviewer, Integration, and Protection agents with human-in-the-loop feedback to ensure alignment with user intent while embedding imperceptible digital watermarks. We formalize the pipeline as a joint optimization objective unifying controllability, semantic alignment, and protection robustness. This work contributes to responsible generative AI by positioning multi-agent architectures as a solution for trustworthy creative workflows with built-in ownership tracking and content traceability.
\end{abstract}


\section{Introduction}

The rapid advancement of generative artificial intelligence has revolutionized content creation across text, image, and video domains. Systems like GPT-4~\cite{gpt4}, DALL-E~\cite{dalle2}, and Stable Diffusion~\cite{stablediffusion} enable unprecedented creative possibilities, transforming how content is produced in professional and commercial contexts. However, two fundamental challenges limit the adoption of these technologies in real-world applications.

First, current generative systems often operate as ``black boxes,'' providing limited controllability over output characteristics beyond initial prompts~\cite{promptdesign}. Users struggle to achieve precise alignment between their creative intent and generated content, particularly for complex, multi-faceted creations requiring specific compositional elements, stylistic choices, or semantic constraints. This controllability gap becomes critical when precision and iterative refinement are essential.

Second, generated content lacks inherent protection mechanisms, creating risks for copyright infringement, unauthorized usage, and content provenance tracking~\cite{diffusionart}. As generative AI outputs become increasingly sophisticated and difficult to distinguish from human-created content, the need for embedded protection and traceability mechanisms grows urgent. Current watermarking approaches typically apply protection as a post-processing step, which can degrade quality and reduce robustness~\cite{stablesignature}.

While multi-agent frameworks have emerged for complex generative tasks~\cite{mulan,animaker}, existing approaches focus primarily on improving generation quality rather than addressing controllability and protection jointly. Recent systems like MetaGPT~\cite{metagpt} and ChatDev~\cite{chatdev} demonstrate effective role-based collaboration for software development, while TradingAgents~\cite{tradingagents} and NovelSeek~\cite{novelseek} apply multi-agent architectures to financial modeling and literature search respectively. However, none integrate protection mechanisms directly into the generation pipeline or provide the fine-grained human-in-the-loop control necessary for creative applications.

To address these gaps, we introduce a multi-agent generative framework that decomposes creation into controllable stages and embeds protection directly into the pipeline. Our key innovation lies in the \textit{joint optimization} of controllability and protection objectives through specialized agent collaboration. The Planner and Reviewer agents provide iterative alignment between user intent and system output, while the Integration and Protection agents ensure coherent composition and imperceptible watermarking for provenance tracking—all within a unified generation loop rather than as separate post-processing steps.

This paper makes three key contributions: (1) We propose a novel multi-agent generative pipeline that integrates protective AI mechanisms directly into controllable content creation; (2) We formalize the framework through a joint optimization objective that unifies planning, generation, review, integration, and protection; (3) We outline a comprehensive experimental evaluation plan with quantitative metrics and ablation studies to validate the framework's effectiveness.

\section{Related Work}

\subsection{Multi-Agent Generative Systems}

Recent work has demonstrated the effectiveness of multi-agent approaches for complex generative tasks. MuLan~\cite{mulan} pioneered the integration of large language model planning with diffusion models for compositional image generation, showing how agent-based decomposition can improve controllability through structured task breakdown. AniMaker~\cite{animaker} extended this concept to video generation, employing specialized agents including Directors for scene planning, Reviewers for quality assessment, and Post-Production agents for final refinement.

Beyond creative generation, multi-agent frameworks have proven successful in software development and specialized domains. MetaGPT~\cite{metagpt} introduced a collaborative framework where agents assume roles such as product managers, architects, and engineers to produce structured software artifacts. Similarly, ChatDev~\cite{chatdev} demonstrated how role-playing agents can iteratively refine code through simulated organizational structures. TradingAgents~\cite{tradingagents} applied multi-agent collaboration to financial modeling, while NovelSeek~\cite{novelseek} employed agent cooperation for literature search and knowledge discovery.

While these systems demonstrate that role-based decomposition improves task performance, existing approaches focus on either generation quality or task-specific optimization. \textbf{None integrate content protection mechanisms into the generation loop itself, nor do they provide the fine-grained human-in-the-loop control necessary for creative applications where iterative refinement and ownership tracking are paramount.}

\subsection{Generative AI and Information Retrieval}

The intersection of generative AI and information retrieval (IR) has produced significant innovations, particularly in retrieval-augmented generation (RAG) systems~\cite{rag,fid}. These approaches combine retrieved knowledge with generative capabilities to produce contextually relevant content. Recent work has explored multimodal RAG systems that generate images, audio, and video content informed by retrieved information~\cite{multimodalrag,videorag}.

Modern IR workflows increasingly involve not just retrieving existing content, but generating new content tailored to user needs~\cite{rethinkingsearch,beir}. This evolution demands systems that maintain the precision and trustworthiness traditionally associated with IR while embracing generative flexibility. Our framework contributes to this intersection by providing controllable generation mechanisms that align with IR community values of relevance, precision, and verifiable provenance.

\subsection{Protective AI and Content Provenance}

The growing sophistication of generative AI has intensified focus on content protection mechanisms. Recent advances in watermarking for diffusion models show promise for embedding imperceptible markers in generated images~\cite{stablesignature,watermarkrecipe}. Chen et al.~\cite{robustwatermark} demonstrated robust watermarking techniques that survive common image transformations while maintaining generation quality through adversarial training.

Beyond watermarking, researchers have explored fingerprinting techniques for model identification and provenance tracking~\cite{artificialfingerprint,pii}. These approaches enable post-hoc detection of AI-generated content and attribution to specific generation systems.

Despite these advances, existing protective mechanisms are typically applied as post-processing steps rather than being integrated into the generation pipeline. This separation can lead to quality degradation and reduced protection robustness. Our framework uniquely addresses this limitation by embedding the Protection agent directly within the multi-agent loop, enabling protection-aware generation that maintains quality while ensuring robust watermark embedding.

\section{Proposed Multi-Agent Framework}

\subsection{Architecture Overview}

\begin{figure}[t]
    \centering
    \includegraphics[width=0.7\textwidth]{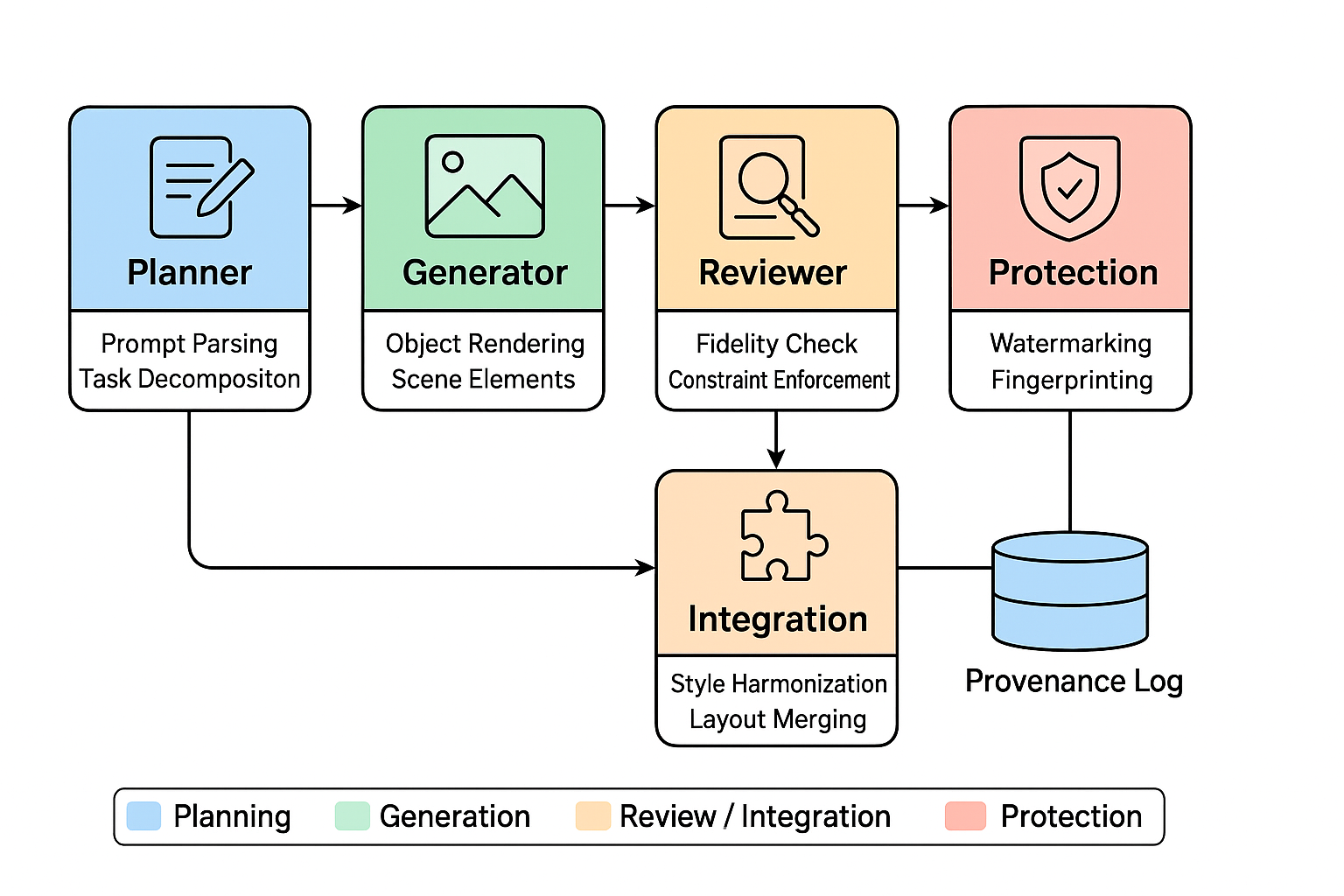}
    \caption{Proposed multi-agent framework.
    The pipeline consists of specialized agents for planning, generation, review, integration, and protection, 
    with iterative human-in-the-loop feedback and provenance logging to ensure both fidelity to user intent 
    and secure content traceability.}
    \label{fig:framework}
\end{figure}

Our proposed framework orchestrates five specialized agents in a sequential pipeline designed to maximize both controllability and protection.Our contribution differs from prior multi-agent systems by (i) integrating a Protection agent within the generation loop—changing both the optimization objective and data flow; (ii) jointly optimizing controllability and watermark robustness in Eq. (10), rather than treating protection as post-hoc; and (iii) exposing protection parameters to the human-in-the-loop controller for traceable creative decisions. Figure~\ref{fig:framework} illustrates the complete workflow, emphasizing the iterative nature and human intervention points that enable fine-grained control over the generation process.

The framework operates on the principle of task decomposition, where complex generative requests are broken down into manageable subtasks handled by domain-specific agents. This modular approach enables targeted optimization of each generation stage while maintaining overall coherence through the Integration agent and ensuring protection through watermark embedding during synthesis rather than post-processing.

\subsection{Agent Roles and Responsibilities}

\textbf{Director/Planner Agent:} Implemented using advanced language models (e.g., GPT-4), this agent serves as the system's strategic coordinator. It analyzes user prompts, decomposes them into specific subtasks, and creates detailed generation plans. For a prompt like ``create a fantasy scene with a red dragon flying above a medieval castle at sunset,'' the Planner would identify key elements (dragon, castle, lighting, composition) and specify generation constraints for each component.

\textbf{Generator Agent:} Responsible for actual content creation, this agent utilizes appropriate generative models based on content type. For images, it employs state-of-the-art diffusion models (e.g., Stable Diffusion XL); for text, transformer-based language models. The Generator receives specific subtask instructions from the Planner and produces initial content iterations.

\textbf{Reviewer/Control Agent:} This agent validates generated content against user intent and enforces quality constraints. Using vision-language models (VLMs) for visual content or specialized evaluation models for text, it assesses alignment with original prompts and identifies areas requiring refinement. The Reviewer can trigger regeneration cycles when outputs fail to meet specified criteria.

\textbf{Integration Agent:} When content involves multiple components or iterations, this agent ensures coherence and consistency across generated elements. It harmonizes style, color palettes, narrative consistency, and other holistic properties that emerge from combining individual generation outputs.

\textbf{Protection Agent:} Operating throughout the generation process, this agent embeds watermarking and fingerprinting mechanisms directly into content creation. Unlike post-processing approaches, integrated protection maintains generation quality while ensuring robust content attribution and ownership tracking.

\subsection{Interactive Control Loop}

A key innovation of our framework is the interactive control mechanism that allows human intervention at any stage of the pipeline. Users can: refine the Planner's decompositions to adjust the overall creative direction; provide targeted feedback to the Generator for specific refinements; override Reviewer assessments when creative intent conflicts with automatic evaluation; request adjustments from the Integration agent to modify style or composition; and configure the Protection agent's parameters to specify watermarking requirements. This human-in-the-loop design directly addresses the controllability challenges inherent in current generative systems while maintaining efficiency through intelligent automation.

\subsection{Mathematical Formulation}

We formalize the multi-agent pipeline as a joint optimization problem that unifies controllability, semantic alignment, coherence, and protection objectives.

\subsubsection{Notation}

Let $P_{\text{text}}$ denote the user's textual prompt describing the desired content. The framework aims to produce a final protected output $I'$ that maximizes alignment with $P_{\text{text}}$ while embedding robust watermarks.

\subsubsection{Planner Agent – Subtask Decomposition}

Given prompt $P_{\text{text}}$, the Planner decomposes it into $k$ subtasks $T = \{T_1, T_2, \ldots, T_k\}$ by maximizing the conditional probability of subtask decomposition given user intent:
\begin{equation}
T^* = \arg\max_{T} \mathcal{P}(T \mid P_{\text{text}}; \theta_p)
\end{equation}
where $\theta_p$ are the Planner model parameters (e.g., GPT-4 weights), and $\mathcal{P}(\cdot)$ denotes probability.

The Planner loss $L_{\text{plan}}$ measures decomposition quality:
\begin{equation}
L_{\text{plan}} = -\log \mathcal{P}(T^* \mid P_{\text{text}}; \theta_p)
\end{equation}

\subsubsection{Generator Agent – Component-wise Synthesis}

For each subtask $T_i$, the Generator produces candidate output $G_i$ using a conditional generative model:
\begin{equation}
G_i = \mathcal{G}(T_i; \theta_g)
\end{equation}
where $\theta_g$ are the Generator model parameters (e.g., Stable Diffusion weights), and $\mathcal{G}$ denotes the generation function.

\subsubsection{Reviewer Agent – Semantic Alignment Scoring}

The Reviewer evaluates alignment between generated component $G_i$ and user prompt $P_{\text{text}}$ using CLIP-based scoring:
\begin{equation}
S_i = \text{CLIP}(G_i, P_{\text{text}})
\end{equation}

The Reviewer enforces a minimum quality threshold $\tau$:
\begin{equation}
G_i \leftarrow \begin{cases}
G_i, & \text{if } S_i \geq \tau \\
\mathcal{G}(T_i; \theta_g), & \text{otherwise (regeneration)}
\end{cases}
\end{equation}

The review loss $L_{\text{rev}}$ penalizes misalignment:
\begin{equation}
L_{\text{rev}} = \sum_{i=1}^{k} \max(0, \tau - S_i)
\end{equation}

\subsubsection{Integration Agent – Coherence Optimization}

The Integration agent merges all generated components into a unified scene $I$. We define the coherence loss as:
\begin{equation}
L_{\text{int}} = \sum_{(i,j) \in \mathcal{N}} \|\Phi(G_i) - \Phi(G_j)\|^2
\end{equation}
where $\Phi(\cdot)$ is a feature extractor for visual embeddings (e.g., VGG or CLIP features), and $\mathcal{N}$ is the set of spatially adjacent component pairs. Minimizing $L_{\text{int}}$ ensures stylistic and spatial harmony.

\subsubsection{Protection Agent – Imperceptible Watermark Embedding}

The Protection agent embeds a secure watermark $W$ into the final image $I$ to produce the protected output $I'$:
\begin{equation}
I' = I + \lambda \cdot W
\end{equation}
where $\lambda$ is the imperceptibility coefficient (typically $\lambda \approx 10^{-3}$ to maintain visual quality).

The protection loss $L_{\text{prot}}$ ensures watermark robustness while maintaining imperceptibility:
\begin{equation}
L_{\text{prot}} = \|I' - I\|^2 + \alpha \cdot \mathcal{R}(W)
\end{equation}
where $\mathcal{R}(W)$ measures watermark recoverability under perturbations, and $\alpha$ balances imperceptibility and robustness.

\subsubsection{Joint Optimization Objective}

The global objective of the framework combines all agent losses:
\begin{equation}
\min_{\theta_g, \theta_p} \left\{ L_{\text{plan}} + L_{\text{rev}} + L_{\text{int}} + L_{\text{prot}} \right\}
\end{equation}

This joint optimization distinguishes our framework from existing approaches by explicitly unifying controllability (via $L_{\text{plan}}$ and $L_{\text{rev}}$), coherence (via $L_{\text{int}}$), and protection (via $L_{\text{prot}}$) within a single objective function.

\subsection{Example Workflow: Text-to-Image Generation}

To illustrate the framework, consider the request: ``Generate an image of a red dragon flying above a medieval castle during a dramatic sunset.'' 

The \textbf{Planner} decomposes the prompt into subtasks: dragon design (pose, color, scale), castle architecture (style, size, positioning), sky composition (sunset colors, cloud formations), and overall scene layout. 

The \textbf{Generator} creates initial versions of each component using targeted prompts derived from this decomposition. 

The \textbf{Reviewer} evaluates intermediate outputs with CLIP-based alignment scoring and may detect issues such as the dragon's positioning obscuring important castle details. 

The \textbf{Integration} agent adjusts the composition to achieve better visual balance while preserving dramatic impact. 

\begin{figure}[H]
    \centering
    \includegraphics[width=0.7\textwidth]{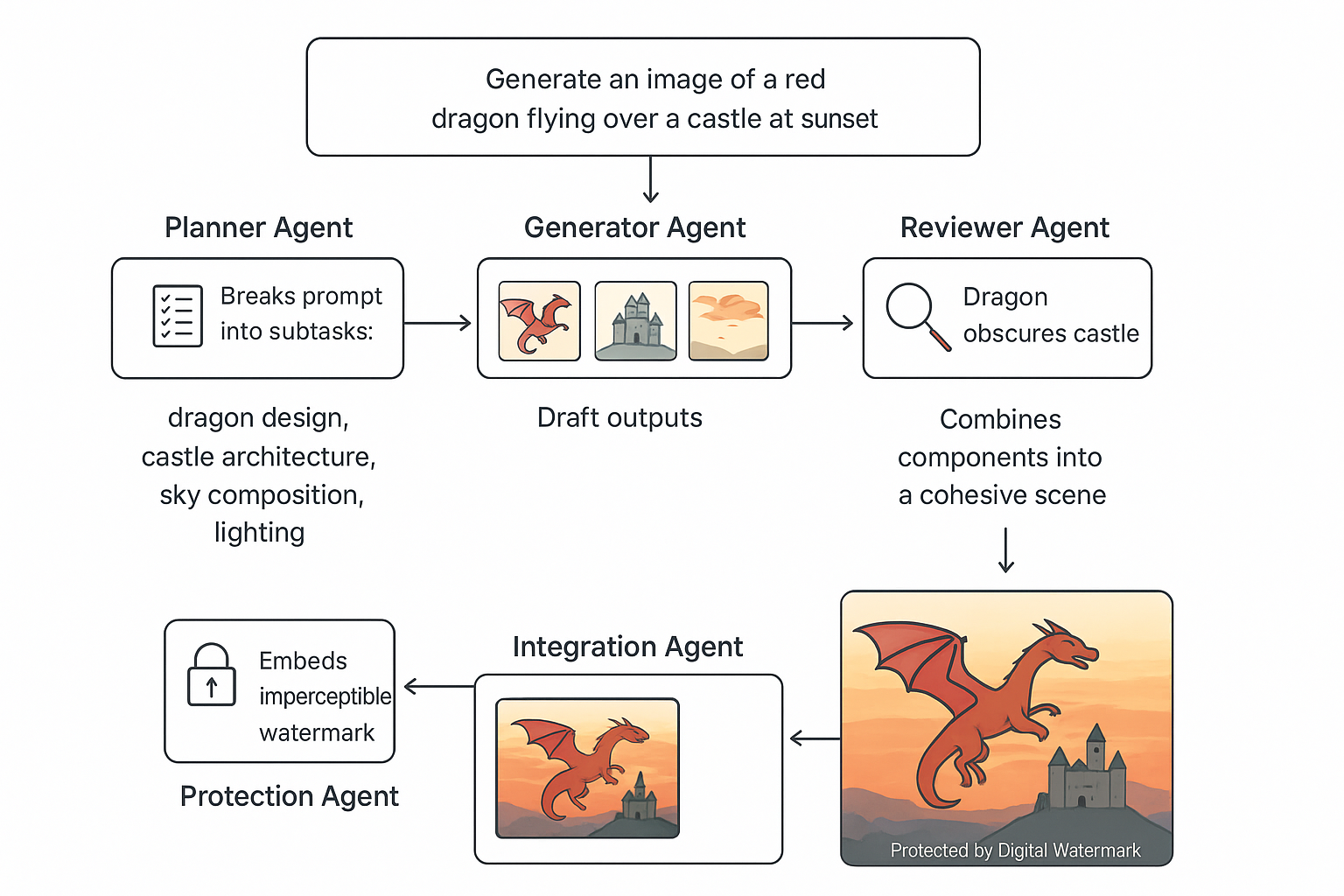}
\caption{Illustrative case study of the multi-agent pipeline using the prompt \textit{``Red dragon flying over a castle at sunset.''} The Planner decomposes subtasks, the Generator synthesizes components, the Reviewer refines alignment, the Integration agent harmonizes layout, and the Protection agent embeds an imperceptible watermark for secure provenance.}
\label{fig:casestudy}
\end{figure}

Finally, the \textbf{Protection} agent embeds an imperceptible watermark into the completed image and records provenance metadata to ensure traceability of the generated content. Figure~\ref{fig:casestudy} visualizes this workflow.

\section{Planned Experimental Evaluation}

To validate the effectiveness of our multi-agent framework, we outline a focused evaluation plan covering controllability, protection robustness, and user experience.

\subsection{Quantitative Metrics}

\textbf{Controllability:} We will measure prompt-output alignment using CLIPScore~\cite{clipscore}, comparing our multi-agent pipeline with single-step generation (e.g., Stable Diffusion without decomposition) across prompts of varying complexity (simple, compositional, and narrative).

\textbf{Generation Quality:} Fréchet Inception Distance (FID) will assess realism and fidelity to ensure that agent-based decomposition and watermarking maintain visual quality relative to baselines.

\textbf{Protection Robustness:} Watermark recovery rate will be evaluated under JPEG compression (quality 50–95), Gaussian noise ($\sigma = 0.01$–$0.05$), cropping (10–30\%), and resizing (50–200\%). Integrated watermarking (during generation) will be compared against post-hoc baselines following Chen et al.~\cite{robustwatermark}.

\subsection{Ablation and User Evaluation}

We will conduct ablations to isolate the contribution of each agent within the framework. In the \textit{No Reviewer} setting, the Reviewer agent will be removed to observe its effect on alignment and iteration efficiency. The \textit{No Integration} variant will generate components independently to evaluate the loss of coherence between elements. The \textit{Post-hoc Protection} configuration will apply watermarking after generation to quantify the robustness improvement gained from in-loop embedding. Finally, the \textit{No Human-in-the-Loop} version will operate in a fully automated mode to assess the importance of interactive control in refining creative outputs.

For user evaluation, a mixed-methods study involving 30–50 participants from creative domains such as digital art, design, and content creation will compare our framework against baseline diffusion systems. The study will measure task completion time, the number of refinement iterations required for user satisfaction, and subjective quality ratings using Likert-scale assessments of controllability, usability, and overall satisfaction. In addition, qualitative feedback from semi-structured interviews will be collected to capture user perceptions of creative control, workflow intuitiveness, and trust in system outputs.

\subsection{Benchmark Datasets}

Evaluation will use COCO~\cite{coco} captions and DrawBench~\cite{drawbench} prompts, supplemented with a curated compositional dataset designed to test multi-object alignment and spatial coherence.

\section{Preliminary Feasibility Analysis}

\subsection{Implementation Approach}

A prototype can be composed from existing models, illustrating immediate feasibility. The Planner uses large language models (e.g., GPT-4) for decomposition, the Generator employs diffusion models such as Stable Diffusion XL, and the Reviewer leverages CLIP-based scoring. Integration is achieved with standard compositing, while the Protection agent builds on watermarking methods for diffusion models~\cite{robustwatermark}. This modular setup demonstrates practical viability without custom training.

\begin{table}[t]
\centering
\caption{Feasibility evidence from prior work, illustrating expected controllability and protection improvements for our integrated pipeline.}
\label{tab:feasibility}
\begin{tabular}{@{}lll@{}}
\toprule
\textbf{Evaluation Aspect} & \textbf{Baseline} & \textbf{Prior Results} \\
\midrule
Controllability & Lower alignment & +20–25\% improvement \\
(CLIPScore) & (single-step) & via decomposition~\cite{mulan} \\
\midrule
Protection robustness & $\sim$70\% recovery & $>$90\% recovery with \\
(JPEG compression) & (post-hoc) & integrated diffusion~\cite{robustwatermark} \\
\midrule
Human-AI interaction & $\sim$4–5 iterations & $\sim$2–3 iterations with \\
(iterations to satisfaction) & (prompt-only) & reviewer feedback~\cite{humanloop} \\
\bottomrule
\end{tabular}
\end{table}
\subsection{Evidence from Prior Work}

Table~\ref{tab:feasibility} summarizes feasibility indicators from related studies. 
MuLan~\cite{mulan} achieved 20–25\% CLIPScore gains from task decomposition, suggesting potential controllability improvement. 
Chen et al.~\cite{robustwatermark} showed integrated watermarking exceeds 90\% recovery under compression (vs. 70\% for post-hoc), validating our protection strategy. 
Human-in-the-loop studies~\cite{humanloop} report fewer refinement iterations with reviewer feedback.

\section{Discussion}

\subsection{Strengths and Advantages}

The proposed multi-agent framework combines controllability, coherence, and protection in a unified loop. Its modular design allows targeted optimization of each stage while maintaining global consistency. 
Watermark embedding during synthesis ensures provenance without quality loss, and the human-in-the-loop design enables precise creative control. 
Unlike frameworks such as MetaGPT~\cite{metagpt} or ChatDev~\cite{chatdev}, our system explicitly optimizes for ownership and provenance—crucial for responsible generative AI.

\subsection{Implications and Applications}

\textbf{Creative Industries:} Artists and designers gain iterative control while preserving authorship through traceable, watermark-protected content.  

\textbf{Content Provenance:} Publishers and media organizations can authenticate and track generative outputs, mitigating misinformation and unauthorized reuse.  

\textbf{Responsible AI:} Integrating protection during generation enforces transparency and intellectual property safeguards, aligning with emerging AI governance frameworks.  

\textbf{Information Retrieval:} Retrieval-augmented variants of our Planner could enable precise, provenance-aware creative retrieval and synthesis.

\subsection{Limitations and Future Work}

Current reliance on third-party APIs and multi-agent orchestration increases computational overhead, though pipeline parallelization could alleviate latency. 
Watermark robustness against targeted removal remains an open challenge~\cite{robustness}. 
Future work will execute the evaluation plan across diverse modalities (image, video, audio), enhance agent coordination efficiency, and strengthen adversarial robustness through hybrid spatial–frequency watermarking.

\section{Conclusion}

This paper presents a novel multi-agent framework that uniquely combines controllable content generation with integrated protection mechanisms, addressing critical challenges in responsible generative AI deployment. Unlike existing multi-agent systems focused solely on generation quality, our approach jointly optimizes controllability and provenance protection through specialized agent collaboration and a unified objective function.

Our key contributions include: (1) a formalized multi-agent pipeline integrating planning, generation, review, integration, and protection within a single optimization framework; (2) human-in-the-loop control mechanisms enabling fine-grained creative refinement; and (3) integration of watermarking during generation rather than post-processing, potentially improving robustness while maintaining quality.

The proposed experimental evaluation plan with quantitative metrics, ablation studies, and user studies provides a rigorous path to validate effectiveness. Feasibility analysis based on existing literature suggests promising potential: prior work indicates 20-25\% controllability improvements through task decomposition and 90\%+ watermark recovery rates with integrated protection approaches.

We call upon the research community to explore the intersection of controllable generation and content protection. As generative AI becomes central to creative workflows, frameworks that balance creativity with ownership protection and user control will prove essential for trustworthy and responsible deployment. The code and evaluation datasets will be made available to facilitate community research in this emerging area.

\section*{Acknowledgments}

We thank the anonymous reviewers for their constructive feedback, which significantly improved this paper. Their suggestions regarding experimental evaluation, mathematical clarity, and positioning relative to existing multi-agent frameworks were invaluable.

\bibliographystyle{plain}

\newpage
\appendix

\section{Algorithmic Workflow}
\label{app:algorithm}

Algorithm~\ref{alg:multiagent} provides a detailed procedural description of the multi-agent generative framework. The algorithm illustrates the sequential coordination between agents and the iterative refinement loop enabled by the Reviewer agent.

\begin{algorithm}[H]
\caption{Multi-Agent Generative Framework}
\label{alg:multiagent}
\begin{algorithmic}[1]
\REQUIRE User prompt $P_{\text{text}}$
\ENSURE Final protected image $I'$
\STATE
\STATE \textbf{// PLANNER: Decompose the prompt into subtasks}
\STATE $T \leftarrow \text{Planner}(P_{\text{text}})$ \quad // $T = \{T_1, T_2, \ldots, T_k\}$
\STATE
\STATE \textbf{// GENERATOR: Generate candidate components}
\FOR{each $T_i$ in $T$}
    \STATE $G_i \leftarrow \text{Generator}(T_i)$ \quad // Generate draft components
\ENDFOR
\STATE
\STATE \textbf{// REVIEWER: Evaluate semantic alignment using CLIP}
\FOR{each $G_i$ in $\{G_1, G_2, \ldots, G_k\}$}
    \STATE $S_i \leftarrow \text{CLIPScore}(G_i, P_{\text{text}})$
    \IF{$S_i < \tau$}
        \STATE $G_i \leftarrow \text{Regenerate}(T_i)$ \quad // Retry if alignment is low
    \ENDIF
\ENDFOR
\STATE
\STATE \textbf{// INTEGRATION: Merge all components into a unified image}
\STATE $I \leftarrow \text{Integration}(\{G_1, G_2, \ldots, G_k\})$
\STATE
\STATE \textbf{// PROTECTION: Embed invisible watermark}
\STATE $I' \leftarrow \text{WatermarkEmbed}(I, \lambda)$
\STATE
\RETURN $I'$
\end{algorithmic}
\end{algorithm}

The algorithm demonstrates how each agent contributes to the overall pipeline while maintaining modularity. The Planner (line 3) establishes the generation strategy, the Generator (lines 5-7) produces content components, the Reviewer (lines 9-14) ensures quality through iterative refinement, the Integration agent (line 16) harmonizes components, and the Protection agent (line 18) embeds watermarks. This structure enables targeted optimization of each stage while preserving end-to-end coherence.

\section{Detailed Agent Specifications}
\label{app:specs}

\subsection{Planner Agent Implementation Details}

The Planner agent utilizes few-shot prompting with GPT-4 or similar large language models. The prompt template includes:
\begin{itemize}
\item \textbf{Task Description:} Explicit instruction to decompose the user prompt into structured subtasks.
\item \textbf{Examples:} 3-5 demonstrations showing prompt decomposition for various complexity levels.
\item \textbf{Output Format:} JSON schema specifying subtask structure (object, attributes, constraints).
\end{itemize}

For compositional prompts requiring spatial relationships, the Planner additionally specifies layout constraints (e.g., ``dragon positioned in upper-right quadrant, castle in lower-center'').

\subsection{Generator Agent Configuration}

The Generator agent employs Stable Diffusion XL with the following configurations:
\begin{itemize}
\item \textbf{Inference Steps:} 50 denoising steps for high-quality generation.
\item \textbf{Guidance Scale:} 7.5 to balance prompt adherence and creative diversity.
\item \textbf{Resolution:} 1024×1024 pixels for detailed component generation.
\item \textbf{Negative Prompting:} Standard quality filters (e.g., ``blurry, distorted, low quality'').
\end{itemize}

\subsection{Reviewer Agent Evaluation Criteria}

The Reviewer agent uses CLIP ViT-L/14 for semantic alignment scoring with threshold $\tau = 0.25$ (empirically determined to balance quality and iteration efficiency). Additional evaluation criteria include:
\begin{itemize}
\item \textbf{Object Presence:} Verification that required objects appear in generated components.
\item \textbf{Attribute Matching:} Color, size, and pose alignment with subtask specifications.
\item \textbf{Artifact Detection:} Identification of visual artifacts (distortions, unnatural elements).
\end{itemize}

\subsection{Integration Agent Composition Strategy}

The Integration agent employs a two-stage process:
\begin{enumerate}
\item \textbf{Spatial Composition:} Components are arranged according to Planner-specified layout using alpha blending and depth-based layering.
\item \textbf{Style Harmonization:} Color palette normalization and lighting consistency adjustments using histogram matching and gradient-domain blending.
\end{enumerate}

\subsection{Protection Agent Watermarking Protocol}

The Protection agent implements a frequency-domain watermarking approach:
\begin{itemize}
\item \textbf{Watermark Generation:} Pseudo-random binary signature derived from content hash and timestamp.
\item \textbf{Embedding:} DCT-based insertion in mid-frequency coefficients with imperceptibility parameter $\lambda = 0.001$.
\item \textbf{Metadata Recording:} JSON provenance log including generation timestamp, model versions, user identifier (hashed), and watermark parameters.
\end{itemize}

The watermark is designed to survive JPEG compression (quality $\geq$ 70), Gaussian noise ($\sigma \leq 0.03$), and cropping ($\leq$ 25\% removal) based on robustness requirements.

\end{document}